\documentclass[article]{IEEEtran}

\usepackage{authblk}
\usepackage{amsfonts,amsmath,amsthm,amssymb,dsfont,bbm}
\usepackage{color}

\ifCLASSINFOpdf 
\usepackage[pdftex,colorlinks, 
           bookmarks=true,%
 pdftitle=Disruptive\ events\ in\ high-density\ high-density\ networks,%
           pdfauthor=P.\ Keeler%
           \-\ B.\ Jahnel%
           \ - \ O.\ Maye%
           \ - \D.\ Aschenbach%
           \-\M. \ Brzozowski,]{hyperref}  
\usepackage[numbers,sort&compress]{natbib} 
\usepackage{hypernat} 
\hypersetup{
   bookmarksnumbered,
   pdfstartview={FitH},
   citecolor={blue},
   linkcolor={red},
   urlcolor=[rgb]{0,0.55,0},
   pdfpagemode={UseOutlines}}
\else

 \usepackage[dvips]{graphicx} 
 \usepackage[numbers,sort&compress]{natbib} 
\fi

\newcommand{\E}{\mathbf{E}}

\newcommand{\V}{\mathbb{V}}

\renewcommand{\P}{\mathbf{P}}

\newcommand{\R}{\mathbb{R}}

\newcommand{\SIR}{\text{SIR}}

\def\D{\Delta}
\def\a{\alpha}
\def\sm{\setminus}

\def\phi{\varphi}

\def\la{\lambda}

\def\de{\delta}

\def\D{\Delta}

\def\P{{\Phi}}

\def\T{\T}

\def\one{\mathms 1}

\def\V|{{\Vert}}

\def\E{\mathbb{E}}

\def\one{\mathbbmss{1}}

\def\ms{\mathsf}

\def\one{\mathbbmss{1}}
\def\P{\mathbb{P}}
\def\R{\mathbb{R}}

\def\threshold{\tau_{\lambda}}
\def\threshconst{\tau}

\newtheorem{thm}{Theorem}
\newtheorem{cor}[thm]{Corollary}

\newtheorem{lem}[thm]{Lemma}
\newtheorem{prop}[thm]{Proposition}
\newtheorem{rem}[thm]{Remark}

\title{
Disruptive events in high-density cellular networks 
}

\author[1]{H.~P.~Keeler~\thanks{keeler@ens.fr}}
\author[2]{B.~Jahnel~\thanks{jahnel@wias-berlin.de}}
\author[3]{O.~Maye~\thanks{maye@ihp-microelectronics.com}}
\author[3]{D.~Aschenbach~\thanks{aschenbach@ihp-microelectronics.com}}
\author[3]{M.~Brzozowski~\thanks{brzozowski@ihp-microelectronics.com}}

\affil[1]{Inria/ENS, Paris, France}
\affil[2]{Weierstrass Institute, Berlin, Germany}
\affil[3]{IHP, Frankfurt (Oder), Germany}

\begin{document}

\maketitle

\begin{abstract}
Stochastic geometry models are used to study wireless networks, particularly cellular phone networks, but most of the research focuses on the typical user, often ignoring atypical events, which can be highly disruptive and of interest to network operators. We examine atypical events when an unexpected large proportion of users are disconnected or connected by proposing a hybrid approach based on ray launching simulation and point process theory. This work is motivated by recent results~\cite{wireless2} using large deviations theory applied to the signal-to-interference ratio. This theory provides a tool for the stochastic analysis of atypical but disruptive events, particularly when the density of transmitters is high. For a section of a European city, we introduce a new stochastic model of a single network cell that uses ray launching data generated with the open source RaLaNS package, giving deterministic path loss values. We collect statistics on the fraction of (dis)connected users in the uplink, and observe that the probability of an unexpected large proportion of disconnected users decreases exponentially when the transmitter density increases. This observation implies that denser networks become more stable in the sense that the probability of the fraction of (dis)connected users deviating from its mean, is exponentially small. We also empirically obtain and illustrate the density of users for network configurations in the disruptive event, which highlights the fact that such bottleneck behaviour not only stems from too many users at the cell boundary, but also from the near-far effect of many users in the immediate vicinity of the base station. We discuss the implications of these findings and  outline possible future research directions. 
\end{abstract}

\begin{IEEEkeywords}
Atypical network configurations, large deviations, ray launching
\end{IEEEkeywords}

\section{Introduction}
To better understand wireless networks, researchers have successfully developed mathematical models of networks based on point processes, where the main performance quantity is the signal-to-interference ratio (SIR). The network models are routinely based on the Poisson point process, which is the most tractable spatial point process~\cite{FnT1,haenggi2012stochastic}, resulting often in closed-form expressions~\cite{sinrmoments,book2018stochastic}. Furthermore,  to any single observer in the network, the network will appear more Poisson in terms of how received signal power values behave randomly, provided there are sufficient random propagation effects such as shadowing multi-path fading~\cite{hextopoi-journal,keeler2014wireless}, even if there is some degree of correlation~\cite{ross2016wireless}. 

Stochastic network models usually share two general characteristics. First, the models largely reflect the experience of a typical user in the network, made formal by Palm calculus, where the focus is often on deriving the probability distribution of the SIR of the typical user. Second, the signal propagation model is a deterministic distance-dependent path loss function coupled with independent and identically distributed (iid) random variables, which represent propagation effects such as shadowing or fading. 

We depart from these standard network model assumptions by offering a new approach, where its novelty stems from tackling the SIR analysis with two methods, one stochastic and one deterministic, coupled together. Our aim is to study atypical or rare events of the network in terms of SIR values. More specifically, we will examine rare events of the single-cell random network model, with the emphasis on analyzing the fraction of users that are (or not) connected  to the central base station. The analysis of such unlikely but disruptive events is often difficult because simulations are expensive for rare events. This work should help network designers not only to quantify the probability of such bottleneck events, but also to determine and efficiently simulate typical user configurations that lead to the disruption behaviour. 

Although we achieve our results through stochastic simulations, we are motivated and guided by recent SIR-based mathematical results by Hirsch, Jahnel, Keeler and Patterson~\cite{wireless2} using large deviations theory, which is a powerful probabilistic tool for studying atypical events in random systems. More precisely, in our case, this theory establishes an asymptotic rate of exponential convergence to zero for the probability of certain atypical events in the limit of high-user densities, which is an increasingly relevant setting, particularly given the rise of so-called ultra-dense networks~\cite{kamel2016ultra}. The use of large deviations theory to study random systems in physics is standard~\cite{El06,touchette2009large}, but a recent handful of papers~\cite{ldpInt,ldpGinibre,wireless1,konig2017gibbsian,huang2013analytical} have also used it to study interference and SIR in wireless networks. 

In addition to studying atypical events, the other key aspect of this work is to replace the standard propagation model, consisting of a path loss function and iid random variables, with path loss values estimated by using ray launching simulations and building data~\cite{AsBrMa17}.  The ray launching is performed with the open source package RaLaNS~\cite{hanel2015ralans} and building data is gathered from publicly available city models~\cite{Berlin3D}.  Our new hybrid approach offers a way to combine stochastic and deterministic simulation techniques, resulting in more realistic models and simulations.

Overall, this paper has two principal aims. First, we wish to demonstrate the hybrid approach for a simple network model, without introducing additional model complexities such as power control or additional network relays. Second, we wish to show the insight gained from recent results based on large deviations theory. We support these results by collecting statistics on the fraction of (dis)connected users. Important empirical observations include that the probability of unlikely events vanishes exponentially fast as the transmitter density increases and users clustered around a base station decrease the overall connectivity (in the uplink) in a network cell.

\begin{figure}
\begin{minipage}{.5\textwidth}
\begin{center}
\centerline{\includegraphics[scale=0.4]{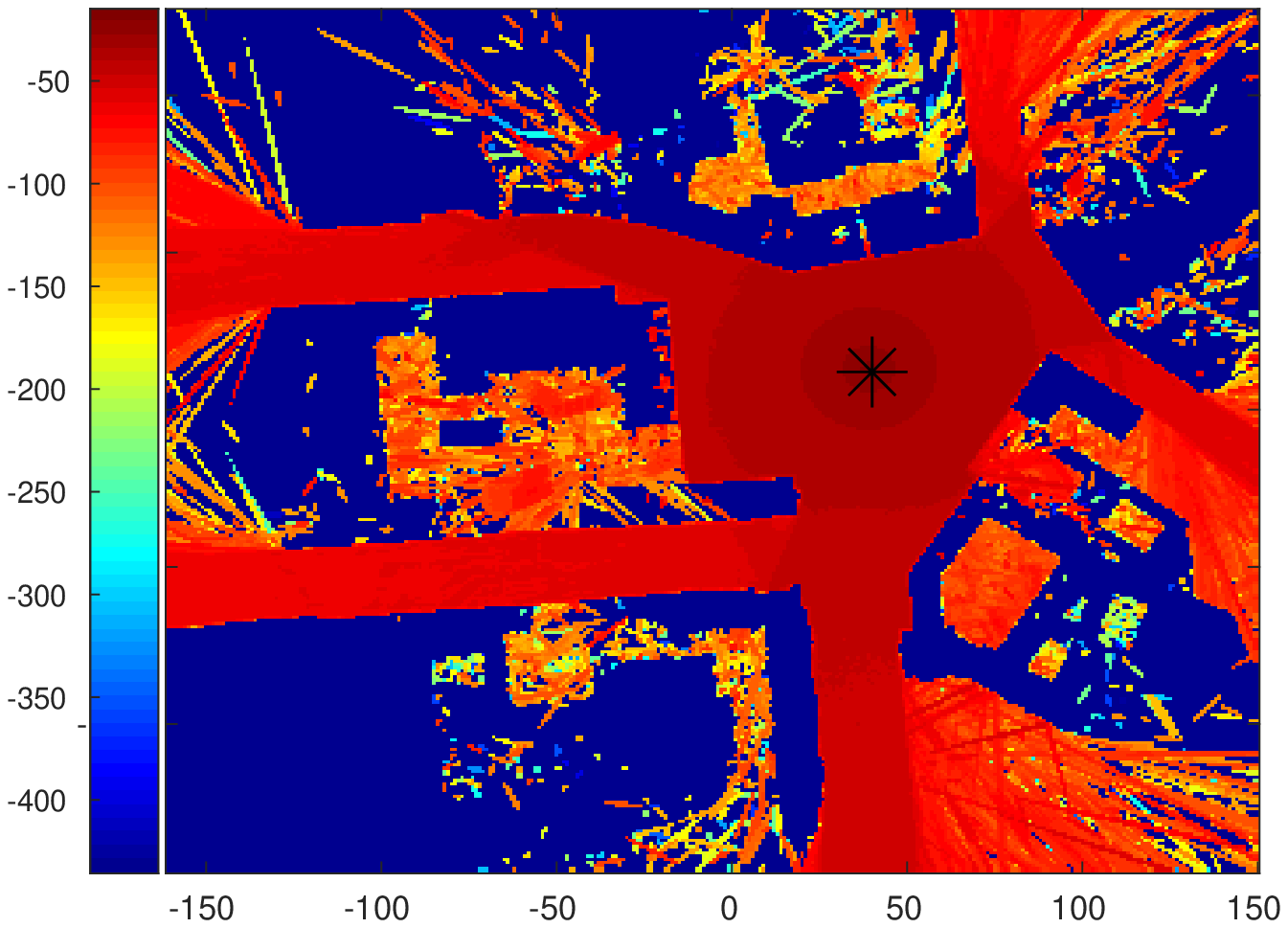}}
\caption{A heat map of path loss values (in decibels) around Hausvogteiplatz in Berlin, estimated using a ray launching method from the open source package RaLaNS~\cite{hanel2015ralans}. The rectangular region has dimensions $311$m by $274$m. The large $\ast$ symbol denotes the simulated signal source, intrepreted here as a base station. \label{Heatmap}}
\end{center}
\end{minipage}
\begin{minipage}{.5\textwidth}
\begin{center}
\centerline{\includegraphics[scale=0.4]{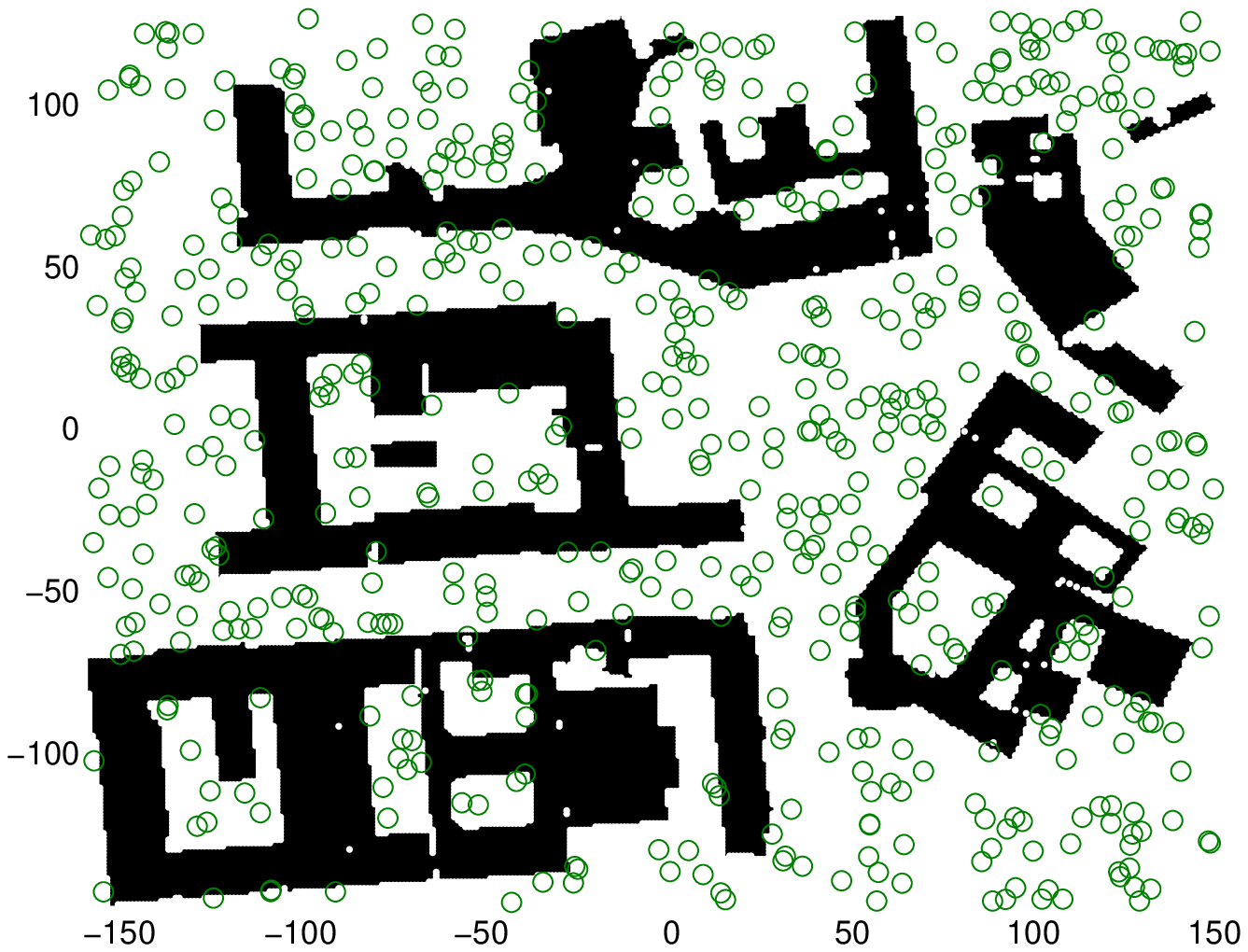}}
\caption{A Poisson point process with density $\la=0.01$ representing users scattered around Hausvogteiplatz in Berlin, away from buildings. \label{Hausvogteiplatz}}
\end{center}
\end{minipage}
\end{figure}
 
\section{Network model}\label{s.model}
We consider a single base station positioned at the center $o$ of some region of a city represented by a bounded and closed set $ W\subset\R^2$. We assume user locations are distributed according to a Poisson point process $X^\la$ with \emph{intensity measure} $\la\mu$, where $\la>0$ is a scaling parameter. In the homogeneous setting, $\mu$ is simply the two-dimensional Lebesgue measure (or area) and $\la$ represents the expected number of users per unit area.

For our single-cell network model, we focus only on the uplink scenario without power control, where messages are directed towards the base station. 
For a single user located at $X_i \in X^\la$, its SIR in relation to the target base station at $o$ is given by 
\begin{equation}\label{SIR}
\SIR'(X_i):=\frac{\ell(|X_i|)}{\sum_{X_k\in X^\la\sm \{X_i\}}\ell(|X_k|)},
\end{equation}
where $|\cdot|$ denotes the Euclidean distance and $\ell$ is the \emph{path loss}, which we assume to be bounded in order for the large deviations results to hold~\cite{wireless2}. The path loss is often assumed to be a function of the form $\ell_\a(s)=\min\{1,s^{-\a}\}$, where $\a>0$ is the \emph{path loss exponent}. This framework also works if definition \eqref{SIR} is generalized to include random fading or shadowing variables and noise terms; see the recent thesis by T{\'o}bi{\'a}s~\cite{tobias2016highly}.
In our case, $\ell$ is given by data provided by the aforementioned ray tracing simulations, which we detail in Section~\ref{ss.ray}.

A direct connection from $X_i$ to the base station can be established successfully if $\SIR'(X_i)\ge \threshold$ for a connectivity parameter $\threshold>0$, known as the \emph{SIR threshold}. When working in the high-density regime, that is, as $\la\rightarrow\infty$, one can safely ignore finite contributions such as the signal term $\ell(|X_i|)$ or finite noise in the denominator of equation (\ref{SIR}), 
and work instead with the quantity
\begin{equation}\label{STIR}
\SIR(X_i):=\frac{\ell(|X_i|)}{\sum_{X_k\in X^\la}\ell(|X_k|) }.
\end{equation}
This is sometimes called the signal-to-total-interference ratio, but we continue to refer to it as the SIR. Note that, in this setting, the user $X_i$ is unable to connect to the base station if 
$$\ell(|X_i|)<\threshold \sum_{X_k\in X^\la}\ell(|X_k|)=\la\threshold L_\la(\ell(|\cdot|)),$$
where 
\begin{equation}\label{measurePoisson}
L_\la:=\la^{-1}\sum_{X_k\in X^\la}\de_{X_k} \, ,
\end{equation}
denotes the \emph{empirical measure} of the Poisson point process $X^\la$. Here, for sets $B\subset \R^2$, the Dirac measure $\de_{x}(B)=1$ if and only if $x\in B$.

Empirical measures are especially amenable for large deviations analysis, as we will outline in Section~\ref{ss.ldt}.
In the limit of high densities, 
with probability one,
$L_\la$ converges (weakly) to $\mu$. In particular, as $\la\rightarrow\infty$, the SIR values $\SIR(X_i)$ tend to zero if $\threshold$ stays fixed, leading to a trivial limiting connectivity model. Hence, in order to guarantee comparability under the high-density limit, we assume the $\SIR$ threshold to be proportional to $\la$ by setting $\la\threshold=\tau$.

Our main quantity of interest is the \emph{proportion of disconnected users}, namely
\begin{equation}\label{measureSIR}
L_\la[\SIR]:=\la^{-1}\sum_{X_i\in X^\la}\one\{\SIR(X_i)< \threshconst\}.
\end{equation}
Put simply, the larger the random scalar quantity $L_\la[\SIR]$, the larger the fraction of disconnected users in the network. It is the distribution of $L_\la[\SIR]$ that we want to study in the limit of many users.

\section{Techniques}

\subsection{Large deviations theory}\label{ss.ldt}
Large deviations theory aims to determine the asymptotic exponential rate of decay of the probabilities of unlikely events in some limiting regime. 
The classic introduction to the theory starts with a collection of iid random variables   $T_1, \dots,T_n $ and their empirical (or sample)  mean
\begin{equation}\label{samplemean}
S_n:= \frac 1 n \sum_{i=1}^n T_i \,.
\end{equation}
If $T_n$ are Gaussian  with mean $m$ and standard deviation  $\sigma$, then the empirical mean's probability density $p(s)$ behaves asymptotically as
\begin{equation}\label{probdens}
p(s)=e^{-n J(s)+o(n)}, 
\end{equation}
where
\[
 \qquad J(s)=(s-m)^2/(2 \sigma^2) , \qquad s\in \R \,,
\]
is  called the \emph{rate function}. For large $n$, the density $p$ concentrates around the minimizer of $J(s)$, which in this case is the mean $m$. Away from the minimizer, the probability density converges to zero exponentially fast with rate $J$. This observation can be captured in the following `in a nutshell' version of a \emph{large deviation principle}
\begin{equation}\label{LDP0}
\lim_{n\rightarrow\infty} \frac{1}{n} \log \P(S_n\approx s)  = -J(s).
\end{equation}

As another example, consider iid random variables distributed according to an exponential distribution ${\ms {E}_m}$ with probability density $p_m(s)=m^{-1}e^{-s/m}$. The probability density of their empirical mean has the same exponential form (\ref{probdens}), and the rate function can be expressed via the \emph{relative entropy} (or Kullback-Leibler divergence), which we denote by $h$. More precisely, the rate function $J$ is given by the relative entropy of an exponential distribution with mean $s>0$ with respect to the initial (or a priori) distribution ${\ms {E}_m}$, namely
\begin{equation}\label{RelEnt}
J(s)= h(\ms {E}_s|\ms {E}_m)=\int p_s(t)\log\frac{p_s(t)}{p_m(t)}{\rm d}t\,.
\end{equation}
In particular, for exponential distributions, 
\begin{align}
J(s)= 
s/m-1-\log(s/m),
\end{align}
which again attains its minimum at $s=m$. The appearance of relative entropies in the rate functions is a universal feature of a large class of stochastic processes, and it will also appear in our results below. 

The above types of large deviations results that focus on empirical means are called \emph{level-1 large deviations}. In so-called \emph{level-2 large deviations}, the empirical mean is replaced by another class of random variable, \emph{empirical measure}, defined as
 \begin{equation} \label{empmeasure}
L_n:= \frac 1 n \sum_{i=1}^n \delta_{T_i}.
 \end{equation}
Essentially, the random quantity $L_n(B)$ represents the random proportion of variables of the stochastic process $T_1,\dots,T_n$ 
taking values in the set $B$. 
 
 Empirical measures contain far more information than empirical means. For example,  $S_n$ is just the expectation of the random measure $L_n$. It is the principal goal of large deviations theory to derive limit statements of the form~\eqref{LDP0} for a great variety of functionals of stochastic processes; see also the standard textbook~\cite{dz98} or, for a lighter introduction, the article~\cite{touchette2009large}.  

Despite the theory of large deviations being mathematically technical at times, it proves useful in giving insight into the properties of random systems deviating away from their expected or typical behaviour. In our case, 
where we replace $n$ with the density $\la$,
it has been recently proven in~\cite[Corollary 1.2.]{wireless2} that the following large deviations result for the proportion of connectable users holds,
\begin{equation}\label{LDP}
\lim_{\la\rightarrow\infty}\frac{1}{\la}\log\P(L_\la[\SIR]>b)=-\inf_{\nu:\,\nu[\SIR]> b }h(\nu|\mu) \,.
\end{equation}
Here, $h$ denotes the relative entropy of measures $\nu$ and $\mu$ on $W$, with a definition analogous to expression~\eqref{RelEnt}. The quantity $\nu[\SIR]$ is defined in the same way as $L_\la[\SIR]$ (see  equation \eqref{measureSIR}), but where the empirical measure $L_\la$ (see equation \eqref{measurePoisson}) is replaced (twice) by the measure $\nu$; see also~\cite[Section 1]{wireless2} for details.

In the large deviations result \eqref{LDP}, if $b$ is sufficiently large, so that the event is unlikely in the sense that $b>\mu[\SIR]$, the rate function $\inf_{\nu:\,\nu[\SIR]> b }h(\nu|\mu)$ is strictly positive. This, in particular, implies exponential decay of the probability for the unlikely event; see~\cite[Corollary 1.3.]{wireless2}. 

Looking beyond the mathematical details, the intuition of the large deviations result (\ref{LDP}) is that events that are functions of the SIR in the uplink, such as the number of (dis)connected users, are highly unlikely to deviate away from their means in the high-density regime. In other words, the probability of such events happening decreases exponentially as the number of users increases. From an operator's perspective, this implies that,
if the networks can handle high-density situations then, in such regimes, atypical events happen very rarely and the system behaves highly predictable. Let us mention that, in this work we focus on the proportion of disconnected users as defined in~\eqref{measureSIR}, but also other functions of SIR can be studied using the same approach.

Returning to the large deviations result (\ref{LDP}), further investigation of the minimizing measure $\nu$ can provide deeper insight into the most likely behaviour of the system conditioned on the unlikely event, via Gibbs conditioning. Additionally, it can be used to derive importance sampling schemes, which allows one to simulate the unlikely events more efficiently, through a change of measure; see for example \cite{isSir,KaKa14}. This is particularly helpful because the unlikely events are (exponentially) rare and therefore expensive to simulate. However, calculating or estimating the rate function is a deterministic optimization problem, which is often challenging, so this approach may only be beneficial for very high user densities. This is one of the principal reasons why we propose and use a novel hybrid simulation approach.

\subsection{Ray launching simulations}\label{ss.ray}
Ray launching is a deterministic simulation method that uses geographical and material information of buildings to estimate how electromagnetic signals propagate in environments such as cities. The method is based on shooting rays from a transmitter location, which are then reflected, diffracted, and scattered through a virtual environment. The physical process of electromagnetic wave propagation is abstracted through simplified mathematical equations. Naturally, the results are always an approximation to reality, where the quality of the approximation depends on the computational effort.

For our ray launching simulations,  we considered a rectangular region or window $W$ (with dimensions $311$m by $274$m) surrounding the small plaza \emph{Hausvogteiplatz}, located in the heart of the German city Berlin. The input data is publicly available building data from the Berlin three-dimensional city model~\cite{Berlin3D}. This data collection provides a CityGML description, which is an open standard format for building models, at a level of detail called LoD2. This level of detail is fine enough to see a simplified three dimensional outer shell of every building, as well as simplified roof shapes. We note that the building height information is also contained in the model, and it can be considered in subsequent work. 

To select a realistic scenario, while at the same time keeping the computational effort at a reasonable level, we took a rectangular sample window of Hausvogteiplatz, as described above, from the total data set. We then manually performed  minor corrections by eliminating redundant points in the polygonal description of the building surfaces, which also ensured that the polygons are planar, producing compatible input data. The resulting corrected CityGML data is part of the supplementary data collection, which is available online~\cite{AsBrMa17}.

To perform the ray tracing, we used a package called RaLaNS~\cite{hanel2015ralans}, which consists of two main components. The first component is a ray launcher that uses CityGML descriptions of buildings, as well as some additional information describing the number and location of wave emitters, to compute a discrete map of signal strengths. This map is a two-dimensional matrix representing the relative received signal power at a certain point at a fixed height. If a simulation is done for multiple heights or there are more than one source, the result is a collection of matrices. The output data of this step are path loss values given as a readable text file, which can then be further processed.  A visualization, namely a heat map, of a typical set of such path loss data is illustrated in Figure~\ref{Heatmap}. 

For completeness, we sketch the second component of RaLaNS, but the results presented here are based on only the ray launcher component of RaLaNS. The second component is a path loss model that plugs into the  network simulation package ns-3~\cite{riley2010ns}. It provides path loss models for a ns-3 simulation by reading the results of its ray launcher simulation whenever the ns-3 simulation asks for a path loss at a given position. While the ray launcher part is computationally quite expensive, downstream ns-3 network simulations still run very efficiently. Different user mobility models or even traffic relay strategies are also possible.

Returning to the ray launching simulation, we assumed that the physical path loss of electromagnetic waves is bidirectional, so, in a given environment, the relative signal loss from one point to another is the same in the opposite direction. This phenomenon is caused by physical laws and is well reflected by ray launching simulation results as long as the spatial resolution, that is, the number of rays, is sufficiently high. In general, the path loss values are estimated by a ray starting at the base station, which is then measured at a user location. 
We assumed that the same path loss applies for signals originating from the user location. 

\subsection{Stochastic simulations}
For the stochastic simulations, we divided the rectangular region $W$ into tiles $w_{i,j}$ of fixed width $\Delta x$ and height $ \Delta y $, so the tiles form a disjoint partition of $W$ with $\cup_{i,j}w_{i,j}=W$. Using the corrected building data, we wrote a script to indicate where buildings exist and do not~\cite{AsBrMa17}. Based on the ray launching data, we set $\Delta x= \Delta y =1 $, and encoded the discrete building data with a discrete intensity measure
\begin{equation}
\mu_{\rm d}(w_{i,j})=\Delta x\Delta y\one\{w_{i,j}\text{ contains no buildings}\} \, ,
\end{equation}
which replaces $\mu$ as the intensity measures of the Poisson point process of users; see Section~\ref{s.model}. This means that any user in the same tile $w_{i,j}$ will have the same path loss value, but the approximation error is negligible. Now simulating the inhomogeneous Poisson point process on $W$ amounts to simulating an independent Poisson random variable for each tile $w_{i,j}$ with parameter $\la\mu_{\rm d}(w_{i,j})$. A typical realization is illustrated in Figure~\ref{Hausvogteiplatz}.

In our simulations, we want to study the effects of varying the user density $\la$. As outlined in Section~\ref{s.model}, we work with the rescaled threshold $\threshconst={\la}\threshold$, which  means that we adjust the SIR threshold $\threshold$ linearly in the density.
The calibration of $\threshconst$ is done in such a way that we can expect a sizeable proportion of connected and disconnected users. 
More precisely, we set $\threshconst$ to be the average path loss divided by the expected path loss under the measure $\mu_{\rm d}$, which is simply the inverse of the total area of the non-building regions, that is, $\threshconst =\mu_{\rm d}(W)^{-1}$.

To study the probability $\P(L_\la[\SIR]>b)$ via simulations, we also need to make a proper choice for $b$. For a fixed value of $\la$, we set 
$b=\E(L_\la[\SIR])(1+\epsilon)$, 
where $\epsilon>0$ is sufficiently large in order to study the deviations away from the mean, but not too large, since otherwise the rare event becomes too difficult to observe.
 For $\E(L_\la[\SIR])$, we just estimate it using  simulations with a fixed value of $\la$.

\bigskip

\section{Results}
The dimensions of our sample window $W$ means its area is approximately $85000{\rm km}^2$, with roughly a third covered by buildings, so even $\la=1$ results in a very large number of users (more than $50000$). We were still able to perform all the stochastic simulations on a standard desktop machine in reasonably fast times. The number of simulations typically ranged from $1000-10000$, taking seconds or minutes. The notable exception was the heat map of bad configurations illustrated in Figure~\ref{AtypicalDensity}, which took a couple of hours due to the large number of simulations. We usually set $\threshconst=-50$ dB, which approximately corresponds to $\mu_{\rm d}(W)^{-1}$, that is, the inverse of the total area of non-building regions our sample window expressed in decibels.

\subsection{Large deviations}
As suggested by the analytical results~\eqref{LDP}, as we increase $\la$, the atypical events $L_\la[\SIR]>b$ become less and less likely, hence we need more and more simulations to properly estimate $\P(L_\la[\SIR]>b)$. The exponential nature of large deviations implies that a good heuristic for the number of simulations is to use a number proportional to $c e^{d \la }$, for constant $c>0$ and $d>0$. For  our results, we generally used the quantity $1000 e^{\lambda}$ rounded to the nearest whole number. Fortunately, we only needed to simulate our network model for density values $\la\leq1$, which we found were sufficiently large for large deviation results to be observed. This is arguably due to the relatively large sample window $W$. 

\medskip
\subsubsection{The rate function}
\begin{figure}
\begin{minipage}{.5\textwidth}
\begin{center}
\centerline{\includegraphics[scale=0.4]{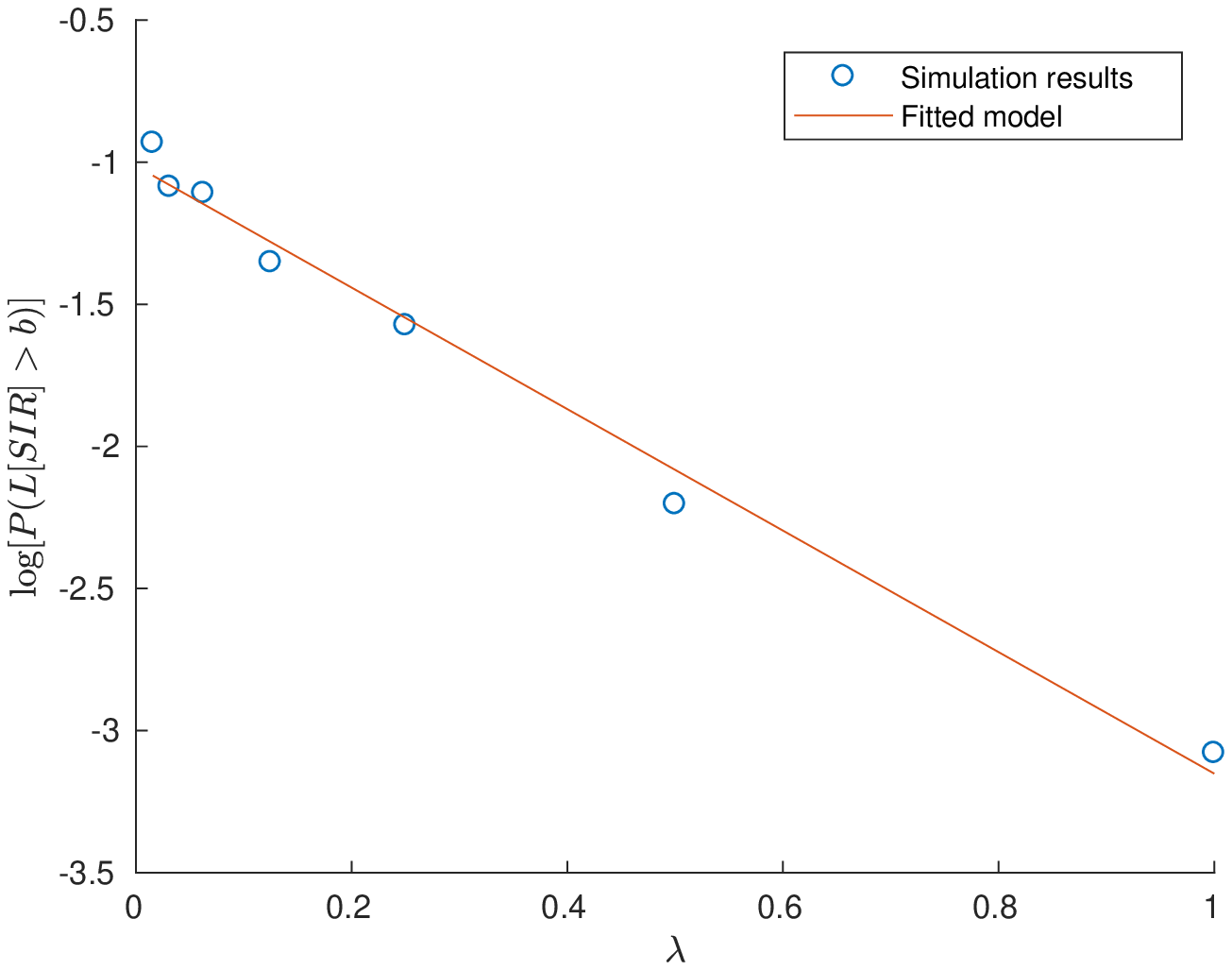}}
\caption{Simulation estimate of $\log\P(L_\la[\SIR]>b)$ with $\threshconst=-50$ dB and $\epsilon=0.01$, fitted to a linear model $\hat{y}_1=p_1\la+p_2$, where $p_1= -2.138 $ and $p_2=-1.014 $. The value of $-p_1$ is an estimate for the rate function. 
\label{LogProbL1}}
\end{center}
\end{minipage}
\begin{minipage}{.5\textwidth}
\begin{center}
\centerline{\includegraphics[scale=0.4]{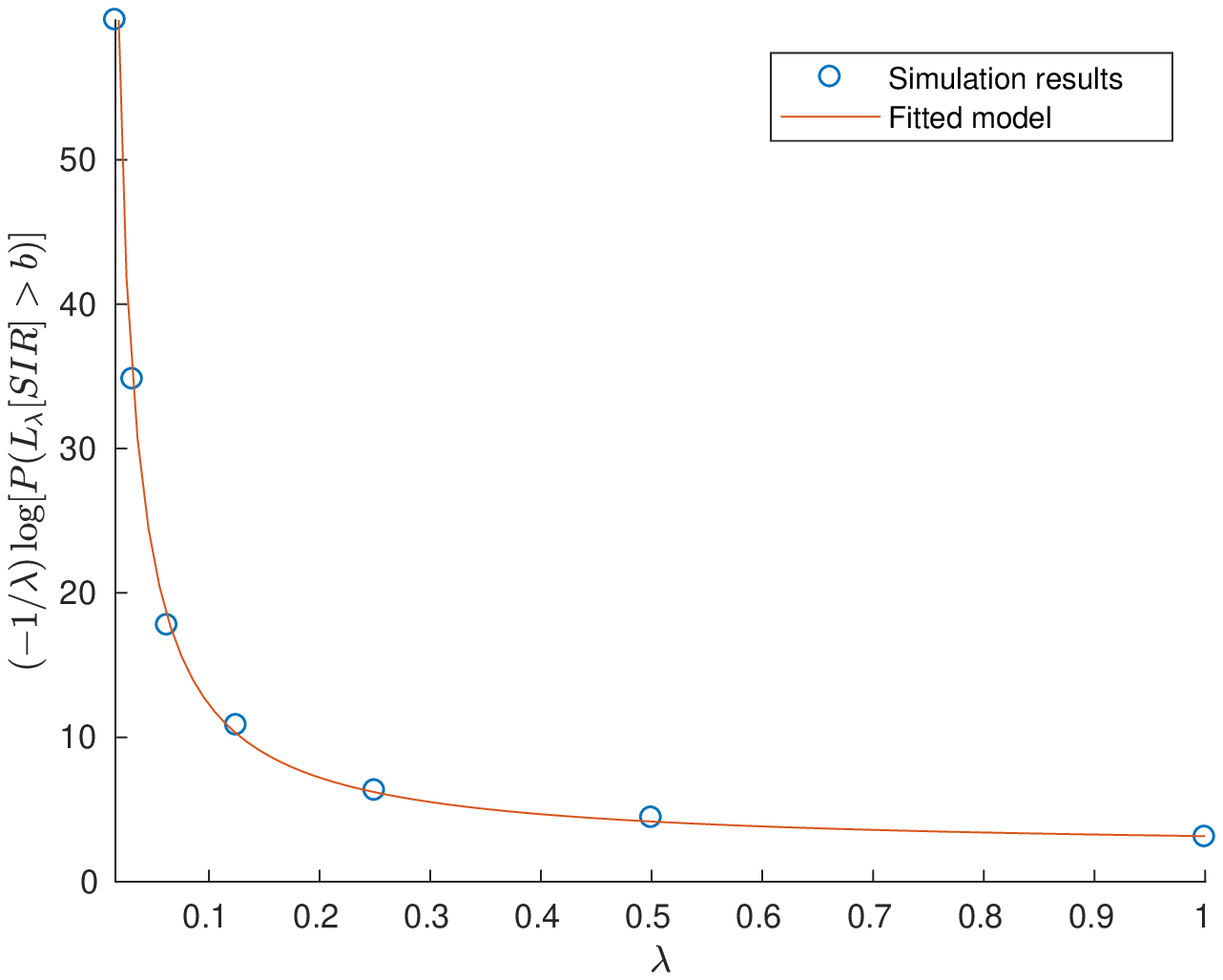}}
\caption{Simulation estimate of $-\log\P(L_\la[\SIR]>b)/\la$ with $\threshconst=-50  $ dB and $\epsilon=0.01$, fitted to a model $\hat{y}_2=p_1+p_2/\la$, where $p_1= -2.138 $ and $p_2=-1.014 $.\label{RateEstimate1}}
\end{center}
\end{minipage}
\end{figure}

We estimated a value of the rate function by simulating our network model, varying the user density $\lambda$, and then collecting statistics on the event $L_\la[\SIR]>b$. This gave us an estimate for $\P(L_\la[\SIR]>b)$, which (after logging) we fitted to a linear model: $\hat{y}_1=p_1\lambda+p_2$. In this linear model $-p_1$ corresponds to the estimate for a value of the rate function.
In Figure~\ref{LogProbL1}, we see that the linear model fits very nicely. Only for small values of $\la$ the simulation results do not fit to the straight line, which suggests that the rate function does not yet dominate in the probability of the deviations. The fitted parameters of the linear models gave us an estimate for the rate function as illustrated in Figure~\ref{RateEstimate1}, where we see it approaches the value $-p_1$. 

\begin{figure}
\begin{minipage}{.5\textwidth}
\begin{center}
\centerline{\includegraphics[scale=0.4]{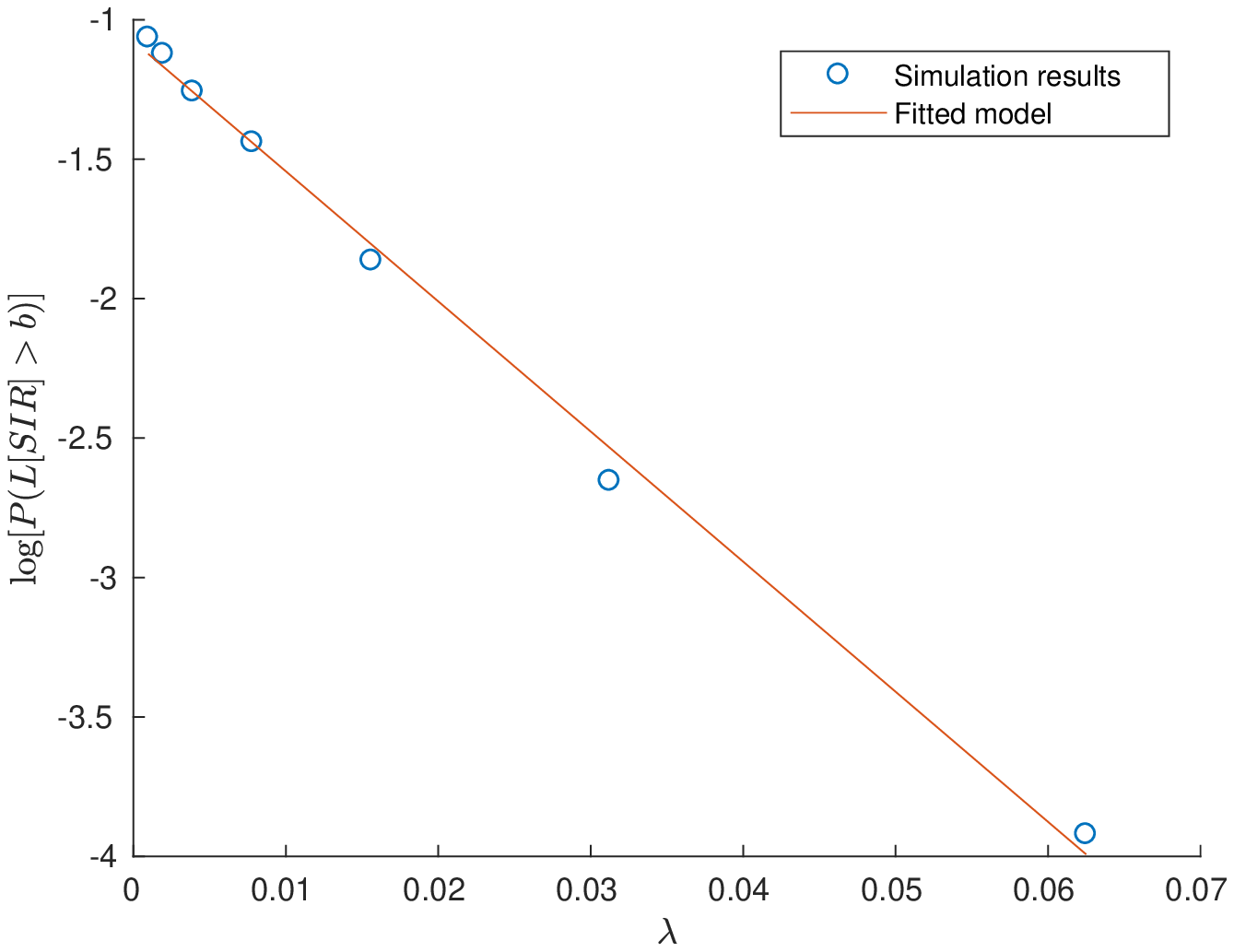}}
\caption{Simulation estimate of $\log\P(L_\la[\SIR]>b)$ with $\threshconst=-50$ dB and $\epsilon=0.05$, fitted to a linear model $\hat{y}_1=p_1\la+p_2$, where $p_1= -46.65 $ and $p_2=-1.077 $. The value of $-p_1$ is an estimate for the rate function. 
\label{LogProbL2}}
\end{center}
\end{minipage}
\begin{minipage}{.5\textwidth}
\begin{center}
\centerline{\includegraphics[scale=0.4]{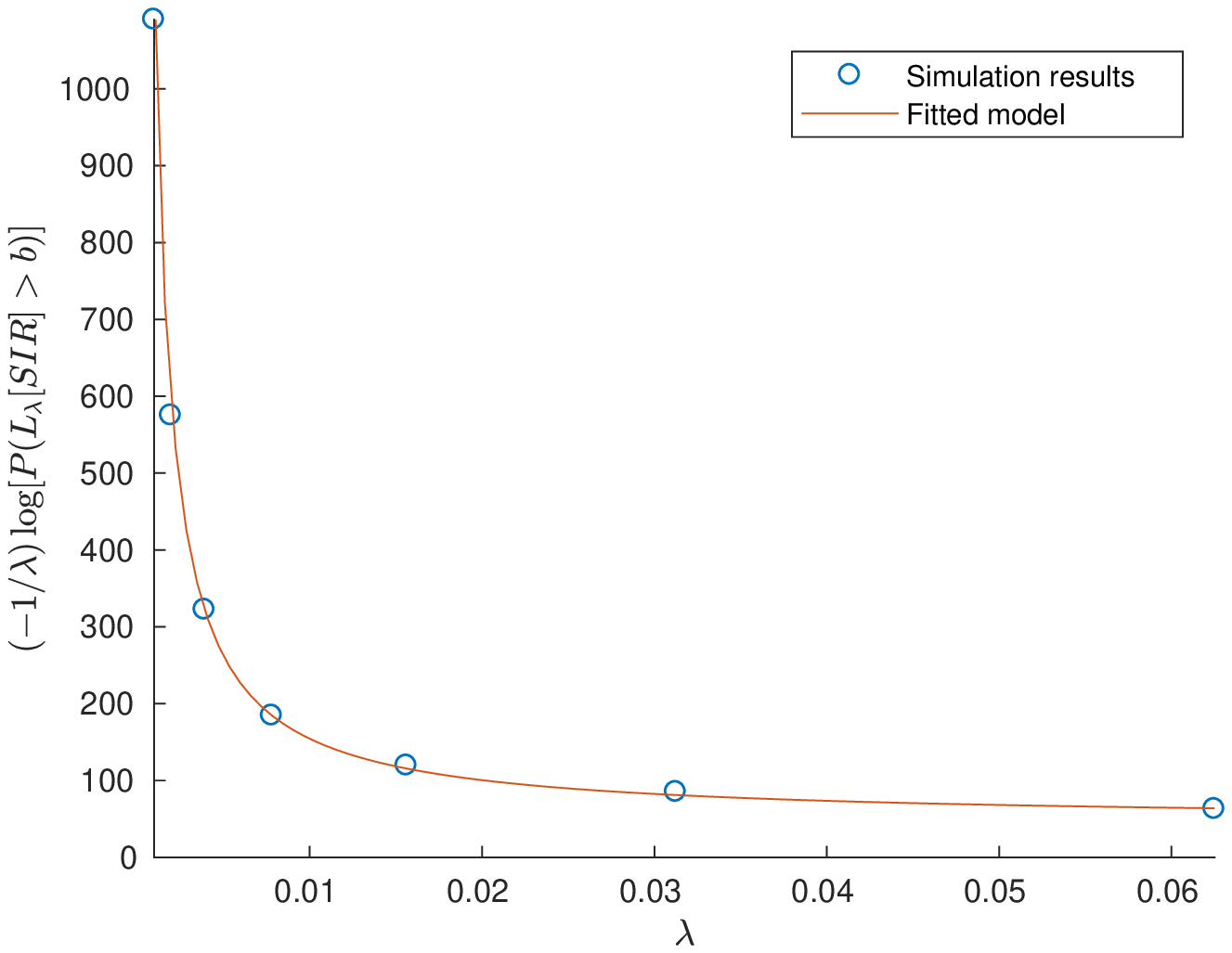}}
\caption{Simulation estimate of $-\log\P(L_\la[\SIR]>b)/\la$ with $\threshconst=-50  $ dB and $\epsilon=0.05$, fitted to a model $\hat{y}_2=-p_1-p_2/\la$,$p_1= -46.65 $ and $p_2=-1.077 $.
\label{RateEstimate2}}
\end{center}
\end{minipage}
\end{figure}

Again, if we increase $b$ (by increasing $\epsilon$), it becomes more difficult to estimate $\P(L_\la[\SIR]>b)$. We see in Figure~\ref{LogProbL2} that the quantity can be estimated for small $\lambda$, but larger $\lambda$ would require more and more simulations. The corresponding estimate for a value of the rate function is given in Figure~\ref{RateEstimate2}.

\medskip
\subsubsection{Bad configurations}
To gain some intuition into the structure of bottleneck configurations, we took an average of user configurations for which a large proportion of users is disconnected. More precisely, we ran many simulations, and then examined the realizations for which $L_\la[\SIR]>b$. Using these atypical realizations, we counted the number of users (that is, Poisson points of $X^\la$) that fall in each tile $w_i\subset W$. For each tile $w_{i,j}$, we then averaged the corresponding ensemble of numbers in order to obtain an estimate for the mean user density for atypical events of a large fraction of disconnected users. 

\begin{figure}
\begin{center}
\centerline{\includegraphics[scale=0.45]{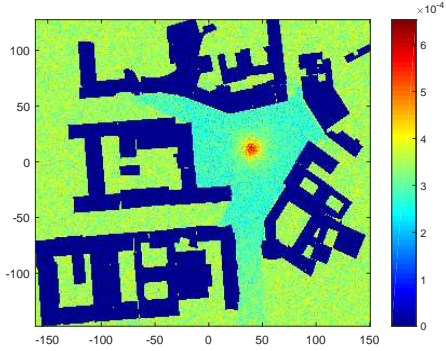}}
\caption{Heat map of bad user configurations with simulation parameters  $\epsilon=0.3$ and $\la=2^{-12}$ (or $\lambda \approx 2.44 \times 10^{-4}$), and $\threshconst=-50 $ (or $\threshold=-13.88
$) dB.  The average user density is estimated for the atypical event of a large fraction of disconnected users. The regions of higher values show where users tend to be located during bad configurations. Naturally, these include regions far or hidden from the target base station. But there is also a region around the base station, due to the so-called near-far effect, where too many users have decreased the SIR values of the remaining network users.  
\label{AtypicalDensity}}
\end{center}
\end{figure}

To perform this step, we reduced the user density down to $\la=2^{-12}$ and increased $\epsilon$. This corresponds to observing larger deviations, which appear less frequently, but, due to the small $\la$, we can balance this by performing more simulations. The results are illustrated as a heat map in Figure~\ref{AtypicalDensity}, which was produced with roughly one million simulations, taking approximately one to two hours on a standard machine. 

Let us explain the intuition behind the heat map in Figure~\ref{AtypicalDensity}. In atypical events, the user density increases in certain areas, showing two different effects. First, more users are located in areas with strong path loss, which are essentially the regions with no line-of-sight connectivity or far from the base station, which is reflected in Figures~\ref{AtypicalDensity} when compared to the path loss values in Figure~\ref{Heatmap}. Second, there is also an increase of users close to the base station, which is due to the \emph{near-far effect} (referred to as the \emph{screening effect} in~\cite{wireless2}), where an accumulation of users increases the interference at the base station. Due to the increased interference, users in certain mid-range areas that would be typically connected to the base station now become disconnected. Figure~\ref{AtypicalDensity} also reveals that, in more abstract terms, it is entropically more favourable to increase the user density in the vicinity of the base station than to place more disconnected users in a much larger area away from the base station. This explains the higher density values around the base station.

Let us further remark that there is also no entropic gain for our process to decrease the density of users in the disruptive event. Indeed, in that event, simulations only show an increase of users in certain areas. Consequently, ignoring the dark blue (building) regions in Figure~\ref{AtypicalDensity}, the low-value regions correspond to the original $\la$ value.

\subsection{Least and most connected configurations}
\begin{figure}
\begin{minipage}{.5\textwidth}
\begin{center}
\centerline{\includegraphics[scale=0.35]{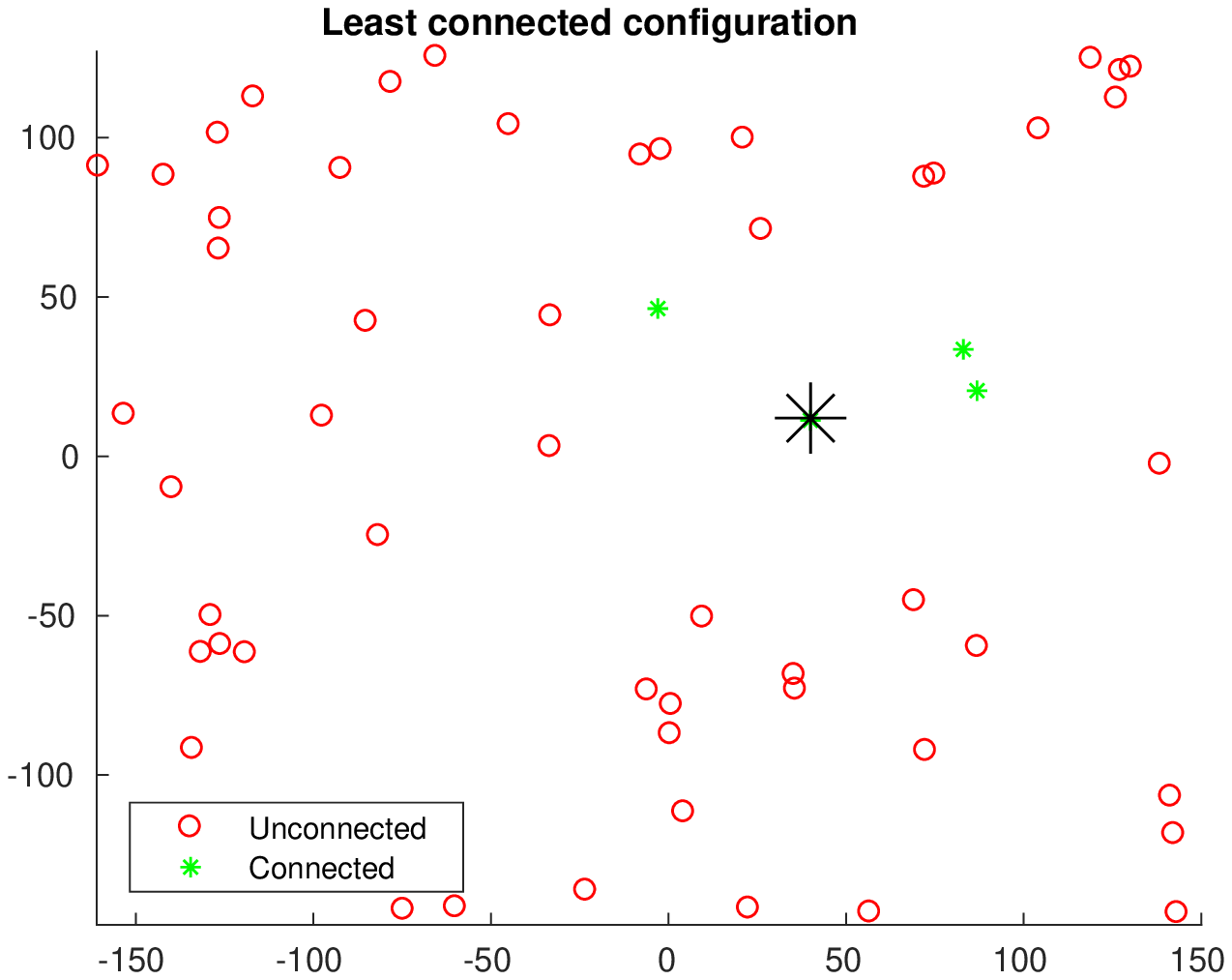}}
\caption{The least connected configuration ($7.02\%
$ connected) of a Poisson point process with density $\la=0.001$ and $\threshconst=-60  $ (or $\threshold=-30$) dB.
\label{Least1}}
\end{center}
\end{minipage}
\begin{minipage}{.5\textwidth}
\begin{center}
\centerline{\includegraphics[scale=0.35]{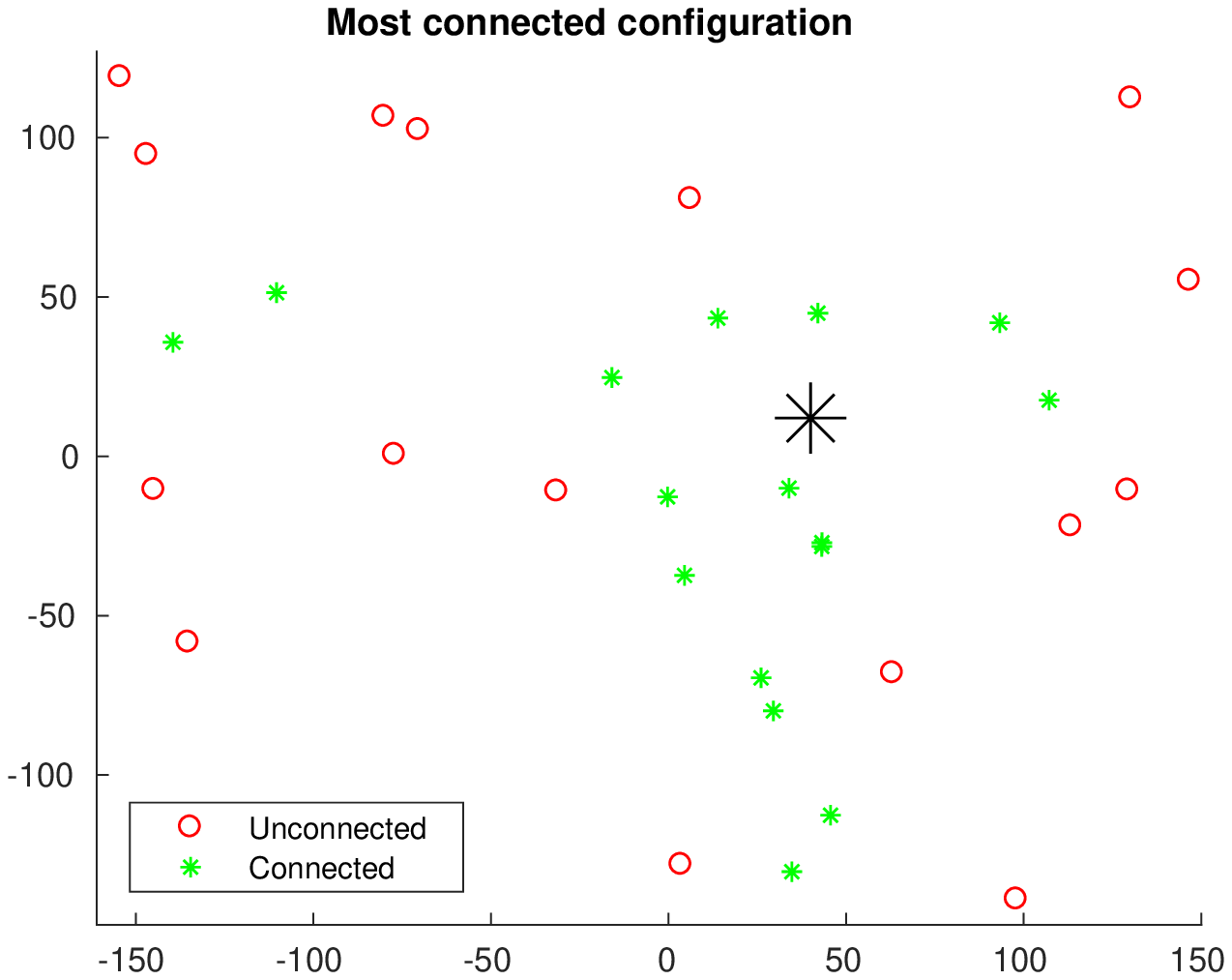}}
\caption{The most connected configuration ($50.00 \%
$ connected) of a Poisson point process with density $\la=0.001$ and $\threshconst=-60  $ (or $\threshold=-30$) dB.\label{Most1}}
\end{center}
\end{minipage}
\end{figure}

\begin{figure}
\begin{minipage}{.5\textwidth}
\begin{center}
\centerline{\includegraphics[scale=0.35]{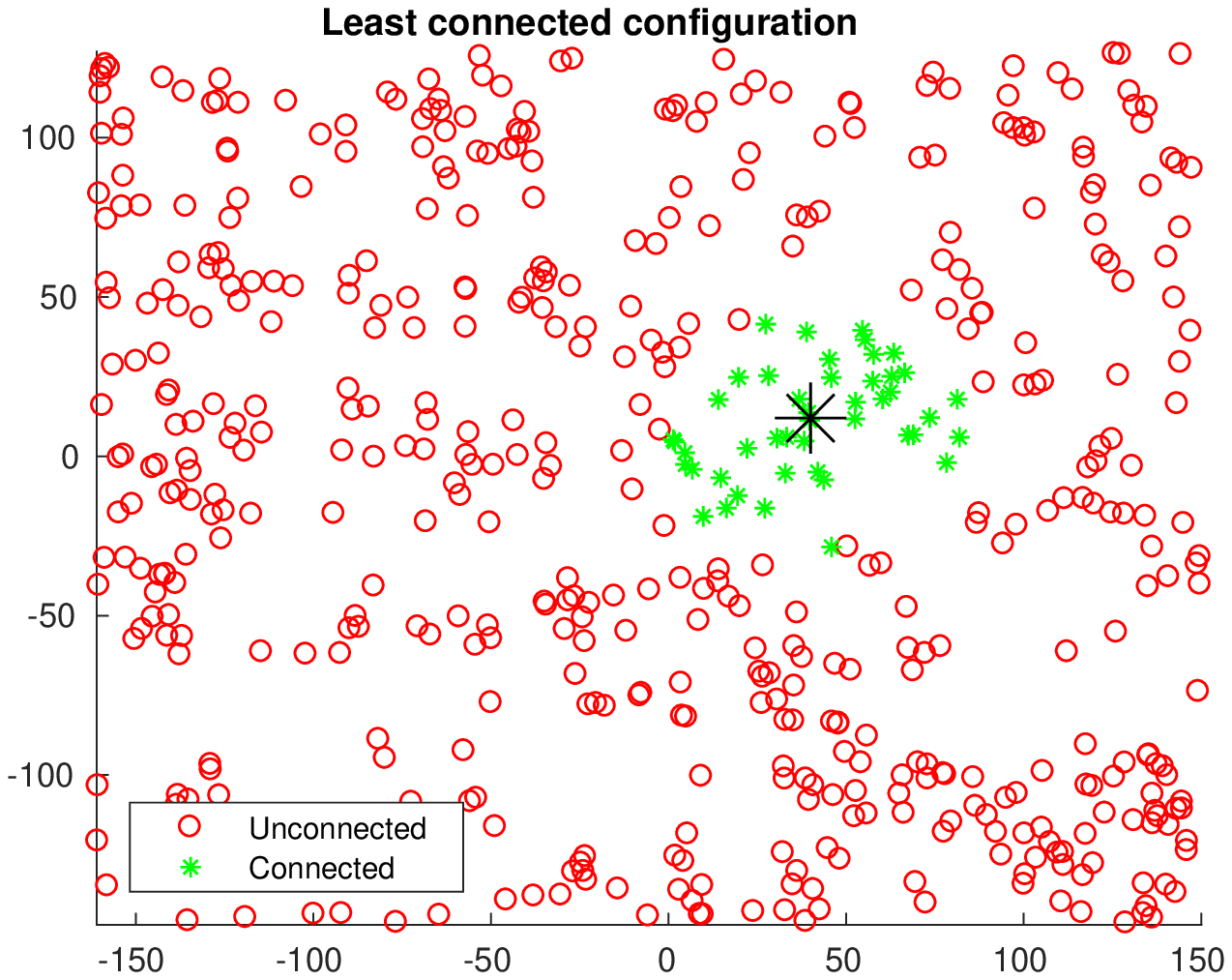}}
\caption{The least connected configuration ($17.99\%
$ connected) of a Poisson point process with density $\la=0.01$ and $\threshconst=-60  $ (or $\threshold=-40$) dB.
\label{Least2}}
\end{center}
\end{minipage}
\begin{minipage}{.5\textwidth}
\begin{center}
\centerline{\includegraphics[scale=0.35]{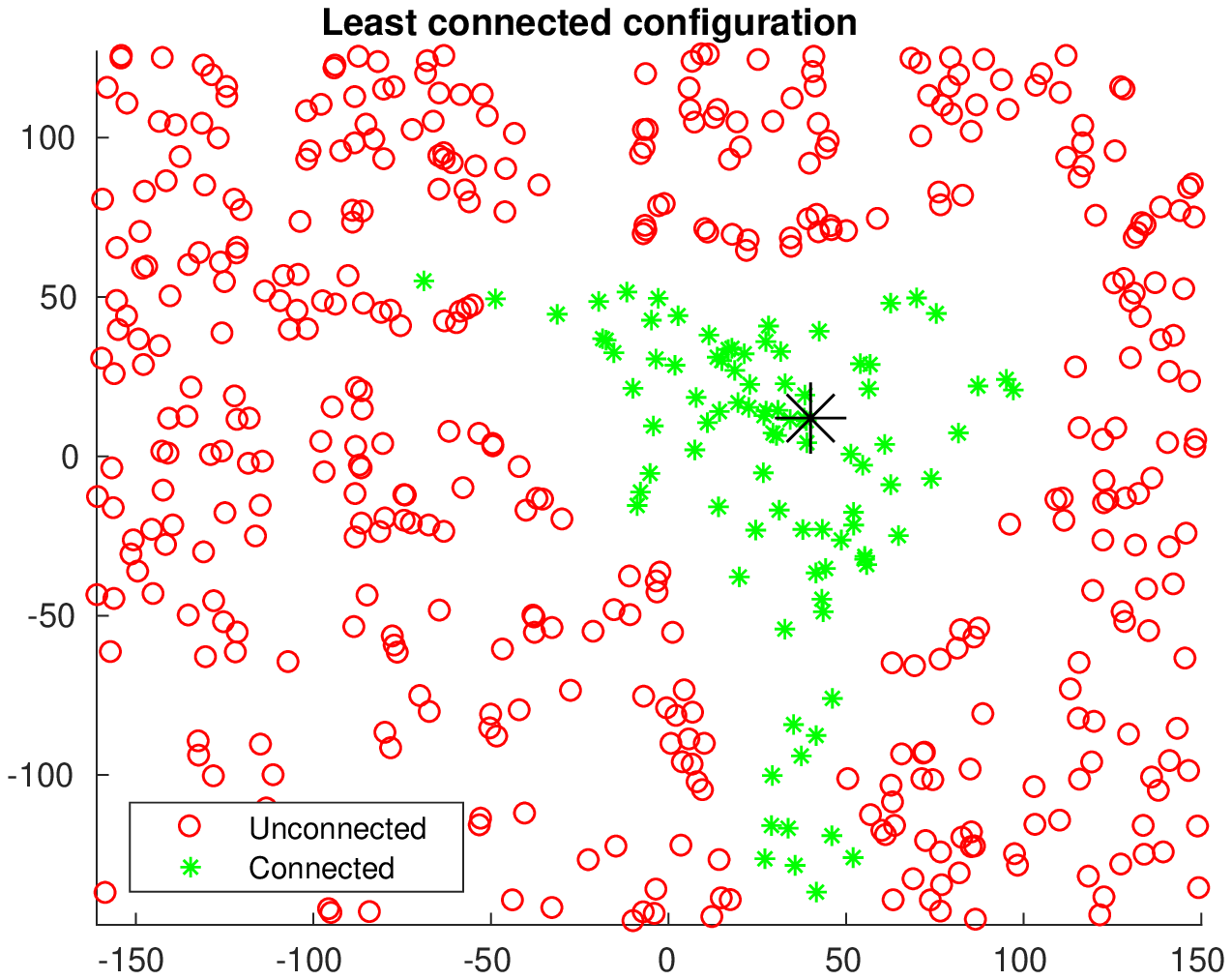}}
\caption{The most connected configuration ($32.06 \%
$ connected) of a Poisson point process with density $\la=0.01$ and $\threshconst=-60  $ (or $\threshold=-40$) dB.\label{Most2}}
\end{center}
\end{minipage}
\end{figure}

In addition to large deviations in our network, we observe how the least and most connected configurations behave. This is done by first performing multiple simulations, and then by examining the configurations with the least and most number of connected users. We observed a lot of variation (or, statistically speaking, variance) in our results for low user density $\la$. For example, in Figures~\ref{Least1} and \ref{Most1}, we see that in the most and least connected configurations (from 10 000 simulations), the fraction of (dis)connected users differs significantly, namely, more than $40\%$, where $\la=0.001$. 
But then for $\la=0.01$, this difference has reduced in Figures~\ref{Least2} and \ref{Most2}. In other words, the difference between the least and most connected configuration decreases as the average number of users per area increases.

%



\section{Discussion and future directions}
In our numerical results, we have seen that the probability of unlikely events, as well as the difference between the most and least connected configuration, decreases as the user density increases.  In terms of connectivity, this observation implies that in dense random networks all configurations appear stochastically (or statistically) the similar.

The  diminishing probability of unlikely events is (potentially) good news in terms of planning for unlikely or rare events, because the probability of such events will vanish in the high-density setting. But of course, dense user configurations will still put strain on networks, even if certain unlikely configurations do not occur. In other words, if a particular network can survive the pressure induced from one high-density configuration, then with high probability, it can survive all other high-density configurations, as stochastically they all behave very similar, with the atypical events almost never occurring. Of course, if a network can withstand any high-density configuration is another question.

We used simulations for the event of an atypical number of disconnected users to build up a portrait of the user density for such configurations. We stress that the heat map in Figure~\ref{AtypicalDensity} does not just reveal where users will experience bad coverage, but also where users will \emph{cause} bad coverage for the remaining network users. The heat map reflects the natural intuition that user-paths with strong path loss are less likely to be connected. More interestingly, it also illustrates the near-far effect, where having too many users clustered around the target base station can be bad for overall network (cell) coverage. This highlights the fact that users should not be allowed to be clustered around base stations. Admittedly, this will not pose a great problem for regular base stations located on isolated rooftops in cities, but the observation is more important for smaller cells located in regions where there are many users. 

In terms of improving the network connectivity, it would be interesting to see the effects of augmenting the network with relays, where, if users cannot connect directly to the base station, they can use an intermediate relay. In fact, a large deviations framework of such a model has already been developed in the paper~\cite{wireless2} that served as the mathematical motivation behind the current work. Using the approaches outlined here, researchers can study the random behaviour of relay-augmented networks. 

The only source of randomness in our network model as well as in the simulations is due to the Poisson point process of users, and not the signals. Consequently, for all our results, the random behaviour is due to the random number and location of users, and not how the signals behave. One possible research direction is incorporating additional layers of randomness, such as fading variables, into the model, and observing the effects. As previously remarked, recent results~\cite{tobias2016highly} show that our model can be generalized to include random variables representing propagation effects such as fading.

\section{Conclusion}
Based on ray launching simulations and recent SIR results using large deviations theory, we have developed a hybrid model approach and studied the connectivity of users in a single-cell scenario of a German city, with the focus being on atypical and disruptive events of a large proportion of users not being connected to the base station.  This work has demonstrated that large deviations theory can be used to gain insight into stochastic network models, thus pointing researchers in directions for further investigation. 

For the average number of users in atypical events, we produced a heat map reflecting bad user configurations in a single cell. It highlights both where users experience bad coverage and where users induce bad coverage for the remaining users. More specifically, it illustrates the near-far effect when a higher than average number of users surround the base station, causing the interference to increase (in the uplink), then users located farther away from the base stations tend to have lower than otherwise expected SIR values.  

A key observation is that in networks there is an interesting trade-off in the high-density regime. From a deterministic perspective, there is increased strain on the network, leading to possible network breakdowns. From a random perspective, however, the probability of random deviations away from the mean decrease exponentially, meaning there is a vanishingly small probability of atypical events occurring, resulting in a type of stochastic stability in the high-density setting. In summary, if a wireless network can be designed to withstand the stress from one high-density  configuration of users, then, in all likelihood, such a network can withstand the stress from all high-density configurations, as stochastically all high-density configurations behave the same.

\section*{Acknowledgements}
This work was supported by the Leibniz program ``Probabilistic methods for mobile ad-hoc networks''. The authors would like to thank Christian Hirsch for stimulating discussions.

\bibliographystyle{IEEEtran}
\bibliography{Platz}

\begin{thebibliography}{10}
\providecommand{\url}[1]{#1}
\csname url@samestyle\endcsname
\providecommand{\newblock}{\relax}
\providecommand{\bibinfo}[2]{#2}
\providecommand{\BIBentrySTDinterwordspacing}{\spaceskip=0pt\relax}
\providecommand{\BIBentryALTinterwordstretchfactor}{4}
\providecommand{\BIBentryALTinterwordspacing}{\spaceskip=\fontdimen2\font plus
\BIBentryALTinterwordstretchfactor\fontdimen3\font minus
  \fontdimen4\font\relax}
\providecommand{\BIBforeignlanguage}[2]{{%
\expandafter\ifx\csname l@#1\endcsname\relax
\typeout{** WARNING: IEEEtran.bst: No hyphenation pattern has been}%
\typeout{** loaded for the language `#1'. Using the pattern for}%
\typeout{** the default language instead.}%
\else
\language=\csname l@#1\endcsname
\fi
#2}}
\providecommand{\BIBdecl}{\relax}
\BIBdecl

\bibitem{wireless2}
C.~Hirsch, B.~Jahnel, P.~Keeler, and R.~Patterson, ``Large deviations in
  relay-augmented wireless networks,'' \emph{To appear in Queueing Systems,
  arXiv:1510.04146}, 2015.

\bibitem{FnT1}
F.~Baccelli and B.~B{\l}aszczyszyn, \emph{Stochastic Geometry and Wireless
  Networks, Volume I --- Theory}, ser. Foundations and Trends in
  Networking.\hskip 1em plus 0.5em minus 0.4em\relax NoW Publishers, 2009.

\bibitem{haenggi2012stochastic}
M.~Haenggi, \emph{Stochastic geometry for wireless networks}.\hskip 1em plus
  0.5em minus 0.4em\relax Cambridge University Press, 2012.

\bibitem{sinrmoments}
B.~B{\l}aszczyszyn and H.~P. Keeler, ``Studying the {SINR} process of the
  typical user in {P}oisson networks by using its factorial moment measures,''
  \emph{IEEE Trans. Inf. Theory}, 2015.

\bibitem{book2018stochastic}
B.~B{\l}aszczyszyn, M.~Haenggi, P.~Keeler, and S.~Mukherjee, \emph{Stochastic
  geometry analysis of cellular networks}.\hskip 1em plus 0.5em minus
  0.4em\relax Cambridge University Press, 2018.

\bibitem{hextopoi-journal}
B.~B{\l}aszczyszyn, M.~K. Karray, and H.~P. Keeler, ``Wireless networks appear
  {P}oissonian due to strong shadowing,'' \emph{IEEE Trans. Wireless Commun.},
  vol.~14, no.~8, pp. 4379--4390, 2015.

\bibitem{keeler2014wireless}
H.~P. Keeler, N.~Ross, and A.~Xia, ``When do wireless network signals appear
  {Poisson}?'' \emph{To appear in Bernoulli, arXiv:1411.3757}, 2017.

\bibitem{ross2016wireless}
N.~Ross and D.~Schuhmacher, ``Wireless network signals with moderately
  correlated shadowing still appear {Poisson},'' \emph{IEEE Transactions on
  Information Theory}, 2016.

\bibitem{kamel2016ultra}
M.~Kamel, W.~Hamouda, and A.~Youssef, ``Ultra-dense networks: A survey,''
  \emph{IEEE Communications Surveys \& Tutorials}, vol.~18, no.~4, pp.
  2522--2545, 2016.

\bibitem{El06}
R.~S. Ellis, \emph{Entropy, large deviations, and statistical mechanics}, ser.
  Classics in Mathematics.\hskip 1em plus 0.5em minus 0.4em\relax
  Springer-Verlag, Berlin, 2006.

\bibitem{touchette2009large}
H.~Touchette, ``The large deviation approach to statistical mechanics,''
  \emph{Physics Reports}, vol. 478, no.~1, pp. 1--69, 2009.

\bibitem{ldpInt}
A.~J. Ganesh and G.~L. Torrisi, ``Large deviations of the interference in a
  wireless communication model,'' \emph{IEEE Trans. Inform. Theory}, vol.~54,
  no.~8, pp. 3505--3517, 2008.

\bibitem{ldpGinibre}
G.~L. Torrisi and E.~Leonardi, ``Large deviations of the interference in the
  {G}inibre network model,'' \emph{Stochastic Systems}, vol.~4, no.~1, pp.
  173--205, 2014.

\bibitem{wireless1}
C.~Hirsch, B.~Jahnel, P.~Keeler, and R.~I. Patterson, ``Large deviation
  principles for connectable receivers in wireless networks,'' \emph{Advances
  in Applied Probability}, vol.~48, no.~4, pp. 1061--1094, 2016.

\bibitem{konig2017gibbsian}
W.~K{\"o}nig and A.~T{\'o}bi{\'a}s, ``A {G}ibbsian model for message routing in
  highly dense multi-hop networks,'' \emph{arXiv:1704.03499}, 2017.

\bibitem{huang2013analytical}
K.~Huang and J.~G. Andrews, ``An analytical framework for multicell cooperation
  via stochastic geometry and large deviations,'' \emph{IEEE transactions on
  information theory}, vol.~59, no.~4, pp. 2501--2516, 2013.

\bibitem{AsBrMa17}
\BIBentryALTinterwordspacing
D.~Aschenbach, M.~Brzozowski, and O.~Maye, ``Disruptive events in high-density
  cellular networks - ray launching data set,'' \emph{Weierstrass Institute
  Berlin}, 2017. [Online]. Available: \url{http://doi.org/10.20347/WIAS.DATA.2}
\BIBentrySTDinterwordspacing

\bibitem{hanel2015ralans}
T.~H{\"a}nel, A.~Bothe, and N.~Aschenbruck, ``{RaLaNS} -- a ray launching based
  propagation loss model for ns-3,'' in \emph{Networked Systems (NetSys), 2015
  International Conference and Workshops on}.\hskip 1em plus 0.5em minus
  0.4em\relax IEEE, 1--7, 2015.

\bibitem{Berlin3D}
``{Berlin 3D - Download Portal.[Online]},''
  \url{http://www.businesslocationcenter.de/en/downloadportal}, accessed:
  2017-11-02.

\bibitem{tobias2016highly}
A.~J. T{\'o}bi{\'a}s, ``Highly dense mobile networks with random fadings,''
  Master's thesis, Technical University of Berlin, \textit{arXiv:1606.06473},
  2016.

\bibitem{dz98}
A.~Dembo and O.~Zeitouni, \emph{Large Deviations Techniques and Applications},
  2nd~ed.\hskip 1em plus 0.5em minus 0.4em\relax Springer, New York, 1998.

\bibitem{isSir}
G.~L. Torrisi and E.~Leonardi, ``Simulating the tail of the interference in a
  {P}oisson network model,'' \emph{IEEE Trans. Inform. Theory}, vol.~59, no.~3,
  pp. 1773--1787, 2013.

\bibitem{KaKa14}
W.~Kang and C.~Kang, ``Large deviations for affine diffusion processes on
  {$\mathbb{R}^m_+\times \mathbb{R}^n$},'' \emph{Stochastic Processes and their
  Applications}, vol. 124, no.~6, pp. 2188 -- 2227, 2014.

\bibitem{riley2010ns}
G.~F. Riley and T.~R. Henderson, ``The ns-3 network simulator,'' \emph{Modeling
  and tools for network simulation}, pp. 15--34, 2010.

\end{thebibliography}
\end{document}